\documentclass[aps,pre,nofootinbib,showpacs,tightenlines,preprint,titlepage,amsmath]{revtex4}

\newcommand{\half}{\mbox{$\textstyle \frac{1}{2}$}}
\newcommand{\cC}{\ensuremath{\mathcal{C}}}
\newcommand{\cP}{\ensuremath{\mathcal{P}}}
\newcommand{\cT}{\ensuremath{\mathcal{T}}}

\begin{document}

\title{Equivalence of a Complex $\cP\cT$-Symmetric Quartic Hamiltonian and a
Hermitian Quartic Hamiltonian with an Anomaly}

\author{Carl~M.~Bender${}^1$, Dorje~C.~Brody${}^2$, Jun-Hua Chen${}^1$,
Hugh~F.~Jones${}^2$, Kimball A. Milton${}^1$,\footnote{Permanent
address: Homer L. Dodge Department of Physics and Astronomy, University of
Oklahoma, Norman OK 73019, USA} and Michael C. Ogilvie${}^1$}

\affiliation{${}^1$Department of Physics, Washington University, St. Louis, MO
63130, USA\\
${}^2$Physics Department, Imperial College, London SW7 2BZ, UK}

\date{\today}

\begin{abstract}
In a recent paper Jones and Mateo used operator techniques to show that the
non-Hermitian
$\cP\cT$-symmetric wrong-sign quartic Hamiltonian $H=\half p^2-gx^4$ has the
same spectrum as the conventional Hermitian Hamiltonian $\tilde H=\half p^2+4g
x^4-\sqrt{2g}\,x$. Here, this equivalence is demonstrated very simply
by means of differential-equation techniques and, more importantly, by means of
functional-integration techniques. It is shown that the linear term in the
Hermitian Hamiltonian is anomalous; that is, this linear term has no classical
analog. The anomaly arises because of the broken parity symmetry of the original
non-Hermitian $\cP\cT$-symmetric Hamiltonian. This anomaly in the Hermitian form
of a $\cP\cT$-symmetric quartic Hamiltonian is unchanged if a harmonic term is 
introduced into $H$. When there is a harmonic term, an immediate physical
consequence of the anomaly is the appearance of bound states; if there were no
anomaly term, there would be no bound states. Possible extensions of this work
to $-\phi^4$ quantum field theory in higher-dimensional space-time are
discussed.
\end{abstract}


\maketitle

\section{Introduction}
\label{s1}

In this paper we consider the quantum system described by the Hamiltonian
\begin{eqnarray}
H=\frac{1}{2m}p^2-gx^4,
\label{e1}
\end{eqnarray}
where $g$ is real and positive. The Hamiltonian (\ref{e1}) is of particular
interest because the corresponding $-\phi^4$ quantum field theory might be a
good model for describing the dynamics of the Higgs sector of the standard
model. This is because the $-\phi^4$ theory is asymptotically free and thus
nontrivial \cite{Sy,BeMi,K}. Furthermore, the one-point Green's function
$\langle0|\phi|0\rangle$ is nonvanishing \cite{FF}.

While the $-x^4$ potential in (\ref{e1}) appears to be unbounded below and the
Hamiltonian $H$ appears to be symmetric under parity reflection, neither of
these properties holds because, as we will explain shortly, the eigenfunctions
of the Hamiltonian are required to vanish exponentially as $|x|\to\infty$ in a
pair of Stokes' wedges in the complex-$x$ plane. These wedges do not include the
real-$x$ axis.

The Hamiltonian (\ref{e1}) is not Hermitian in the conventional Dirac sense,
where Hermitian conjugation is defined as combined matrix transposition and
complex conjugation. Nevertheless, the eigenvalues $E_n$ are all real, positive,
and discrete. This is because $H$ possesses an unbroken $\cP\cT$ symmetry
\cite{B1,B2,R1}, which means that $H$ and its eigenstates $\psi_n(x)$ are
invariant under space-time reflection. Here, $\cP$ denotes spatial reflection
$p\to-p$ and $x\to-x$, and $\cT$ denotes time reversal $p\to-p$, $x\to x$, and
$i\to-i$.

Even though $H$ is not Dirac Hermitian, it is possible to construct a state
space having a positive inner product with respect to which $H$ is Hermitian and
in this Hilbert space, time evolution is unitary (probability conserving)
\cite{B2}. This positive inner product involves $\cC\cP\cT$-conjugation, where
$\cC$ is a linear operator whose square is unity and which commutes with the
Hamiltonian. (These mathematical properties are reminiscent of the charge
conjugation operator in particle physics.) Given the operator $\cC$, we can
construct the {\it positive\/} operator $e^Q=\cC\cP$, which can in turn be
used to construct by means of a similarity transformation an equivalent
Hamiltonian ${\tilde H}\equiv e^{-Q/2}He^{Q/2}$, where $\tilde H$ is Dirac
Hermitian \cite{M}.

Many $\cP\cT$-symmetric quantum-mechanical Hamiltonians have been studied in the
recent literature \cite{R4}. However, the Hamiltonian (\ref{e1}) is especially
interesting because, unlike the $\cP\cT$-symmetric Hamiltonian $\half p^2+ix^3$,
for example, the boundary conditions on the eigenfunctions are not imposed on
the real-$x$ axis. Rather, these boundary conditions hold in the interiors of
wedges in the complex-$x$ plane. To identify these wedges we use a WKB
approximation to determine the possible asymptotic behaviors of the
eigenfunctions. For large $|x|$ the possible exponential behaviors of the
solutions to the Schr\"odinger equation are given by
\begin{equation}
\psi_n(x)\sim e^{\pm i\sqrt{2mg}x^3/3}\quad(|x|\to\infty).
\label{e2}
\end{equation}
This result shows that the eigenfunctions are purely oscillatory along six lines
separated by $60^\circ$ angles. In particular, the wave functions are
oscillatory on the positive and negative real-$x$ axes. These six lines are the
boundaries of the six $60^\circ$ Stokes' wedges in which the solutions in
(\ref{e2}) grow or decay exponentially.

The eigenfunctions for the Hamiltonian (\ref{e1}) are required to decay
exponentially in the interiors of a pair of Stokes' wedges in the lower-half
$x$-plane. These wedges, which are symmetrically placed with respect to the
imaginary axis, lie below the positive and negative real-$x$ axes with the
upper edges of the wedges lying on the real axis.

Note that under space reflection $x\to-x$, the original two-wedge domain changes
to the interior of a {\it different\/} pair of $60^\circ$ wedges that lie in the
{\it upper}-half plane. Therefore, while $H$ in (\ref{e1}) may appear to be
parity symmetric, it obeys parity-violating boundary conditions. One can also
understand the parity violation of $H$ in (\ref{e1}) in a different way. In
Ref.~\cite{ABB} it is shown that the real spectrum of $H$ in (\ref{e1}) can be
obtained without having to impose boundary conditions in the complex plane. The
procedure is simply to require that the $-x^4$ potential be reflectionless.
Thus, an incoming plane wave from $x=-\infty$ propagates past the potential and
becomes an outgoing plane wave at $x=+\infty$. This configuration is not parity
invariant because under parity reflection, the incoming plane wave at $x=-\infty
$ becomes an outgoing plane wave and the outgoing plane wave at $x=\infty$
becomes an incoming plane wave. In short, the right-going flow of probability
current becomes a left-going flow. (Of course, this configuration is invariant
under combined  $\cP$ and $\cT$ reflection.)

This violation of parity symmetry occurs only at the quantum level. At
the classical level, the equations of motion are clearly parity symmetric:
\begin{equation}
\dot x(t)=\frac{1}{m}p(t),\qquad \dot p(t)=4gx^3(t).
\label{e3}
\end{equation}

In general, it is difficult to construct the operator $\cC$, and in the past for
nontrivial models this operator has only been determined perturbatively
\cite{papers}. However, recently Jones and Mateo used perturbative operator
methods to construct $C$ in closed form for the Hamiltonian in (\ref{e1})
\cite{JM}. They then found the equivalent Hermitian Hamiltonian $\tilde H$,
whose potential has a positive quartic term and also a linear term. (Using
Rayleigh-Schr\"odinger perturbation theory, Buslaev and Grecchi had already
discovered this equivalent Hamiltonian much earlier \cite{BG}.)

The purpose of this paper is to examine the connection between the $\cP
\cT$-symmetric non-Hermitian Hamiltonian (\ref{e1}) and the equivalent Hermitian
Hamiltonian $\tilde H$, whose potential has a positive quartic term and a linear
term. In Sec.~\ref{s2} we demonstrate the equivalence between these two
Hamiltonians simply and directly by transforming the Schr\"odinger equation for
the former Hamiltonian into the Schr\"odinger equation for the latter
Hamiltonian. We show that the linear term in $\tilde H$ is proportional to
$\hbar$ and is a therefore quantum anomaly. This anomaly arises because the
boundary conditions on the $\cP\cT$-symmetric Hamiltonian violate parity. As
explained above, this parity violation is not a feature of the classical
equations of motion. In Sec.~\ref{s2.5} we generalize the Hamiltonian H in 
(\ref{e1}) to include a harmonic (quadratic) term and show that the equivalent
Hermitian Hamiltonian has  an anomaly of exactly the same form. Furthermore,
we show that an immediate consequence of the anomaly is the appearance of
bound states.

The Schr\"odinger-equation approach of Sec.~\ref{s2} does not readily generalize
to quantum field theory, so in Sec.~\ref{s3} we demonstrate the equivalence
between $H$ and $\tilde H$ by using path-integration techniques in which we
treat $H$ as defining a quantum field theory in one-dimensional space-time.
We conclude in Sec.~\ref{s4} by indicating how path-integral techniques might be
extended and used to identify the equivalent Hermitian Lagrangian for a
$-\phi^4$ quantum field theory in higher space-time dimensions.

\section{Differential-Equation Derivation of the Anomaly}
\label{s2}

\begin{figure*}[t!]
\vspace{2.1in}
\includegraphics{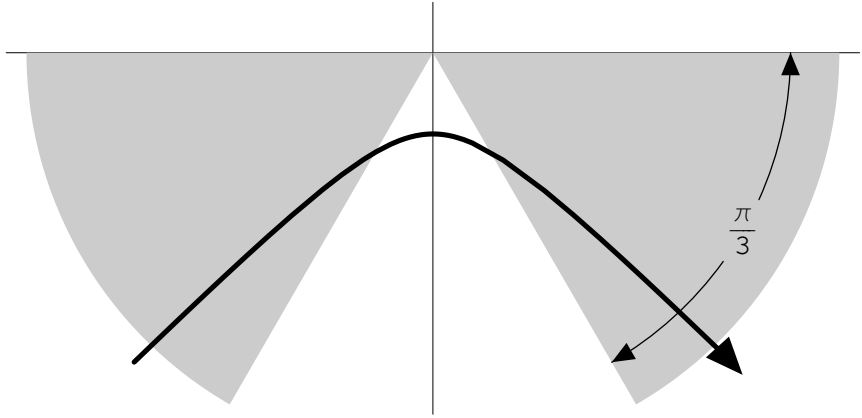}
\caption{Stokes' wedges in the lower-half complex-$x$ plane for the
Schr\"odinger equation (\ref{ee1}) arising from the Hamiltonian $H$ in
(\ref{e1}). The eigenfunctions of $H$ decay exponentially as $|x|\to\infty$
inside these wedges. Also shown is the contour in (\ref{ee2}).}
\label{f1}
\end{figure*}

In this section we use straightforward differential-equation techniques to
establish the equivalence of $H$ and $\tilde H$. Consider the one-dimensional
Schr\"odinger eigenvalue problem
\begin{eqnarray}
-\frac{\hbar^2}{2m}\psi''(x)-gx^4\psi(x)=E\psi(x)
\label{ee1}
\end{eqnarray}
associated with the non-Hermitian Hamiltonian (\ref{e1}), where the boundary
conditions on $\psi(x)$ are that as $\lim_{|x|\to\infty}\psi(x)=0$ if
$-\frac{\pi}{3}<{\rm arg}\,x<0$ and $-\pi<{\rm arg}\,x<-\frac{2\pi}{3}$. These
boundary conditions do not include the real-$x$ axis and they require that the
differential equation (\ref{ee1}) be solved along a contour whose ends lie in
the above wedges in the complex-$x$ plane (see Fig.~\ref{f1}). We will use here
the same complex contour that Jones and Mateo employed in their operator
analysis of the Hamiltonian (\ref{e1}) \cite{JM}:
\begin{eqnarray}
x=-2iL\sqrt{1+iy/L},
\label{ee2}
\end{eqnarray}
where $y$ runs from $-\infty$ to $\infty$ along the real axis.
This contour is acceptable because as $y\to\pm\infty$, ${\rm arg}\,x$ approaches
$-45^\circ$ and $-135^\circ$, so the contour lies inside the Stokes' wedges.
In (\ref{ee2}) $L$ is an arbitrary positive constant having dimensions of
length. In terms of the parameters of $H$ in (\ref{e1}) the fundamental unit of
length is $[\hbar^2/(mg)]^{1/6}$. Thus, 
\begin{eqnarray}
L=\lambda\left(\frac{\hbar^2}{mg}\right)^{1/6},
\label{ee3}
\end{eqnarray}
where $\lambda$ is an arbitrary positive dimensionless constant. When we change
the independent variable in (\ref{ee1}) from $x$ to $y$ according to
(\ref{ee2}), the Schr\"odinger equation (\ref{ee1}) becomes
\begin{eqnarray}
-\frac{\hbar^2}{2m}\left(1+\frac{iy}{L}\right)\phi''(y)-\frac{i\hbar^2}{4Lm}
\phi'(y)-16gL^4\left(1+\frac{iy}{L} \right)^2\phi(y)=E\phi(y).
\label{ee5}
\end{eqnarray}

Next, we perform a Fourier transform of (\ref{ee5}). We define
\begin{eqnarray}
\tilde f(p)\equiv\int_{-\infty}^{\infty} dy\,e^{-iyp/\hbar}f(y),
\label{ee6}
\end{eqnarray}
so that the Fourier transform of $f'(y)$ is $ip\tilde f(p)/\hbar$ and the
Fourier transform of $yf(y)$ is $i\hbar\tilde f'(p)$. Then, the Fourier
transform of the Schr\"odinger equation (\ref{ee5}) is
\begin{eqnarray}
\frac{1}{2m}\left(1-\frac{\hbar}{L}\frac{d}{dp}\right)p^2\tilde\phi(p)
+\frac{\hbar}{4Lm}p\tilde\phi(p) -16gL^4\left(1-\frac{\hbar}{L}\frac{d}{dp}
\right)^2\tilde\phi(p)=E\tilde\phi(p).
\label{ee7}
\end{eqnarray}
Expanding and simplifying this equation, we obtain
\begin{eqnarray}
-16gL^2\hbar^2\tilde\phi''(p)+\left(-\frac{\hbar p^2}{2mL}+32gL^3\hbar\right)
\tilde\phi'(p)+\left(\frac{p^2}{2m}-\frac{3p\hbar}{4mL}-16gL^4\right)
\tilde\phi(p)=E\tilde\phi(p).
\label{ee8}
\end{eqnarray}
[Note that the variable $p$ used here is not the same as the variable $p$ used
in (\ref{e1}). Here, considered as an operator, $p$ represents $-i\hbar\frac{d}
{dy}$, whereas in (\ref{e1}) $p$ represents $-i\hbar\frac{d}{dx}$.]

This equation is not a Schr\"odinger equation because there is a one-derivative
term. However, we can eliminate this term by performing a simple transformation:
\begin{eqnarray}
\tilde\phi(p)=e^{Q(p)/2}\Phi(p).
\label{ee9}
\end{eqnarray}
The condition on $Q(p)$ for which the equation satisfied by $\Phi(p)$ has no
one-derivative term is a first-order differential equation whose solution is
\begin{eqnarray}
Q(p)=\frac{2L}{\hbar}p-\frac{1}{96gmL^3\hbar}p^3.
\label{ee10}
\end{eqnarray}
It is interesting that $e^{Q(p)}$ is precisely the operator found in
Ref.~\cite{JM}.\footnote{Note that $e^{Q(p)}$ is {\it not\/} the $\cC\cP$
operator that could, in principle, be used in a similarity transformation to
produce the Hermitian Hamiltonian from the non-Hermitian $\cP\cT$-symmetric
Hamiltonian (\ref{e1}). One can only use $e^{Q(p)}$ to transform the
non-Hermitian Hamiltonian $H$ to Hermitian form if $H$ has first been written in
terms of the real variable $y$.} Substituting this expression for $Q$ gives the
Schr\"odinger equation satisfied by $\Phi(p)$:
\begin{eqnarray}
-16gL^2\hbar^2\Phi''(p)+\left(-\frac{\hbar p}{4mL}+\frac{p^4}{256gm^2
L^4}\right)\Phi(p)=E\Phi(p).
\label{ee11}
\end{eqnarray}

Finally, we make the scaling substitution
\begin{eqnarray}
p=zL\sqrt{32mg},
\label{ee12}
\end{eqnarray}
to replace the $p$ variable, which has units of momentum, by $z$, which is a
coordinate variable having units of length. The resulting eigenvalue equation,
posed on the real-$z$ axis, is 
\begin{eqnarray}
-\frac{\hbar^2}{2m}\Phi''(z)+\left(-\hbar\sqrt{\frac{2g}{m}}\,z+4gz^4\right)
\Phi(z)=E\Phi(z).
\label{ee13}
\end{eqnarray}
We emphasize that while $z$ has dimensions of length, it is not a conventional
coordinate variable because it is odd under the discrete transformation of time
reversal.

Observe that the eigenvalue problem (\ref{ee13}) is similar in structure to that
in (\ref{ee1}). [Equation (\ref{ee13}) is not {\it dual\/} to (\ref{ee1})
because it is still weakly coupled.] However, the potential has acquired a
linear term, and since this linear term is proportional to $\hbar$, we may
regard this term as a quantum anomaly. The linear term has no classical analog
because the classical equations of motion are parity symmetric. The breaking of
parity symmetry occurs at large values of $x$ where the boundary conditions on
the wave function $\psi(x)$ are imposed. Because we have taken a Fourier
transform to obtain the Schr\"odinger equation (\ref{ee13}), this parity anomaly
now manifests itself at small values of $z$.

The Hamiltonian $\tilde H$ for which (\ref{ee13}) is the eigenvalue problem is
\begin{eqnarray}
\tilde H=\frac{\tilde p^2}{2m}-\hbar\sqrt{\frac{2g}{m}}\,z+4gz^4.
\label{ee14}
\end{eqnarray}
This Hamiltonian is Hermitian in the Dirac sense and is bounded below on the
real-$z$ axis. Furthermore, it is also $\cP\cT$-symmetric. This is because at
every stage in the sequence of transformations above, $\cP\cT$ symmetry is
preserved. However, while $z$ and $\tilde p$ are canonically conjugate operators
satisfying $[z,\tilde p]=i$, the new variable $z$ behaves like a momentum rather
than a coordinate variable because $z$ changes sign under time reversal.

\section{Bound States -- A Direct Physical Consequence of the Anomaly}
\label{s2.5}
If we generalize the Hamiltonian (\ref{e1}) to include a harmonic term in
the potential,
\begin{eqnarray}
H=\frac{1}{2m}p^2+\frac{\mu^2}{2}x^2-gx^4,
\label{x1}
\end{eqnarray}
then the same differential-equation analysis used in Sec.~\ref{s2}
straightforwardly yields the following equivalent Hermitian Hamiltonian:
\begin{eqnarray}
\tilde H=\frac{\tilde p^2}{2m}-\hbar\sqrt{\frac{2g}{m}}\,z+4g\left(z^2-\frac{
\mu^2}{8g}\right)^2.
\label{x2}
\end{eqnarray}
This result is given in Refs.~\cite{BG,JM}. Observe that for these more
general Hamiltonians the form of the linear anomaly term remains unchanged
from that in (\ref{ee13}).

In an earlier paper \cite{BOUND} it was shown that the Hamiltonian
(\ref{x1}) exhibits bounds states. In a particle physics model a bound state
is defined as a state having a negative binding energy. In the context of a
quantum-mechanical model we define bound states as follows: Let the energy
levels of the Hamiltonian be $E_n$ ($n=0,1,2,\ldots$). The renormalized mass
is the mass gap; that is, $M=E_1-E_0$. The higher excitations must also be
measured relative to the vacuum energy: $E_n-E_0$ ($n=2,3,4,\ldots$). We say
that the $n$th higher excitation is a bound state if the binding energy
\begin{equation}
B_n\equiv E_n-E_0-nM
\label{x3}
\end{equation}
is negative. If $B_n$ is positive, then we regard the state as unbound
because this state can decay into $n$ 1-particle states of mass $M$
in the presence of an external field.

It was observed numerically in Ref.~\cite{BOUND} that for small positive values
of $g$ the first few states of $H$ in (\ref{x1}) are bound. As $g$ increases,
the number of bound states decreases until, when $g/\mu^3$ is larger than the
critical value $0.0465$, there are no bound states. In this paper a rough
heuristic argument was given to explain why there is such a critical value.
This argument is difficult to formulate because the non-Hermitian Hamiltonian
is evaluated for $x$ in the complex plane. When $x$ is complex, one cannot use
order relationships such as $>$ or $<$, which only apply to real numbers. 

However, now that we have established that $H$ in (\ref{x1}) has the same
spectrum as the Hermitian Hamiltonian in (\ref{x2}), it is easy to understand
the appearance of bound states. Furthermore, as we will now show, the bound
states are a direct consequence of the linear anomaly term. To probe the
influence of the anomaly, let us generalize (\ref{x2}) by inserting a
parameter $\epsilon$ that measures the strength of the anomaly term:
\begin{eqnarray}
\tilde H=\frac{\tilde p^2}{2}-\epsilon\hbar\sqrt{2g}\,z+
4g\left(z^2-\frac{1}{8g}\right)^2,
\label{x4}
\end{eqnarray}
where for simplicity we have set $m=\mu=\hbar=1$.

\begin{figure}[b!]
\vspace{3.8in}
\includegraphics{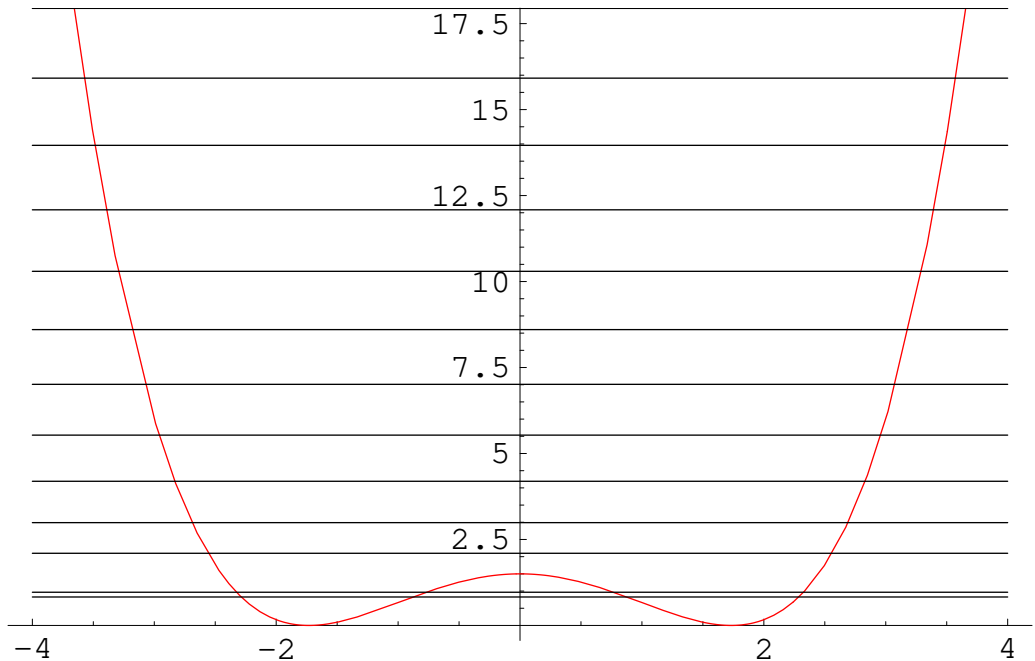}
\caption{Potential of the Hermitian Hamiltonian (\ref{x4}) plotted as a function
of the real variable $z$ for the case $\epsilon=0$ and $g=0.046$. The energy
levels are indicated by horizontal lines. Because $\epsilon=0$ there is no
anomaly, and the double-well potential is symmetric. Therefore, the mass gap is
very small and, as a consequence, there are no bound states at all. The
occurrence of bound states can be attributed to the anomaly.}
\label{f2}
\end{figure}

If we take $\epsilon=0$, then there is no anomaly term and the potential is a
{\it symmetric\/} double well. The mass gap for a double well is exponentially
small because it is a result of the tunneling between the wells. Thus, the
renormalized mass $M$ is very small. Therefore, $B_n$ in (\ref{x3}) is
positive and there are no bound states. In Fig.~\ref{f2} we display the
double-well potential and the first several states of the system for the
case $g=0.046$ and $\epsilon=0$. Note that the lowest two states have a
very small splitting.

\begin{figure}[b!]
\vspace{3.8in}
\includegraphics{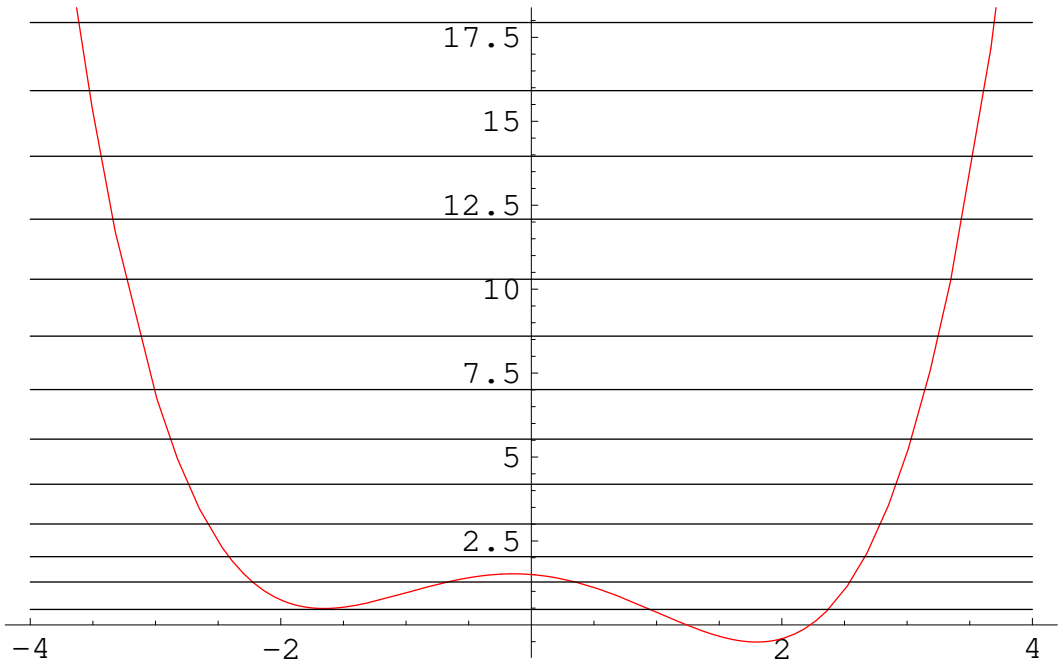}
\caption{Asymmetric potential well plotted as a function of the real variable
$z$ for the Hermitian Hamiltonian (\ref{x4}) with $\epsilon=1$ and $g=0.046$.
The energy levels are indicated by horizontal lines. There is one bound state.}
\label{f3}
\end{figure}

If $\epsilon=1$, the double-well potential is asymmetric and the lowest two
states are not approximately degenerate. As a result, bound states can occur
near the bottom of the potential well. The higher-energy states eventually
become unbound because, as we know from WKB theory, in a quartic well the $n$th
energy level grows like $n^{4/3}$ for large $n$. As $g$ becomes large, the
number of bound states becomes smaller because the 
depth of the double well decreases. For sufficiently large $g$ there are
no bound states. In Fig.~\ref{f3} we display the potential for $\epsilon=1$
for $g=0.046$. For this value of $g$ there is only one bound state.

\begin{figure}[b!]
\vspace{3.8in}
\includegraphics{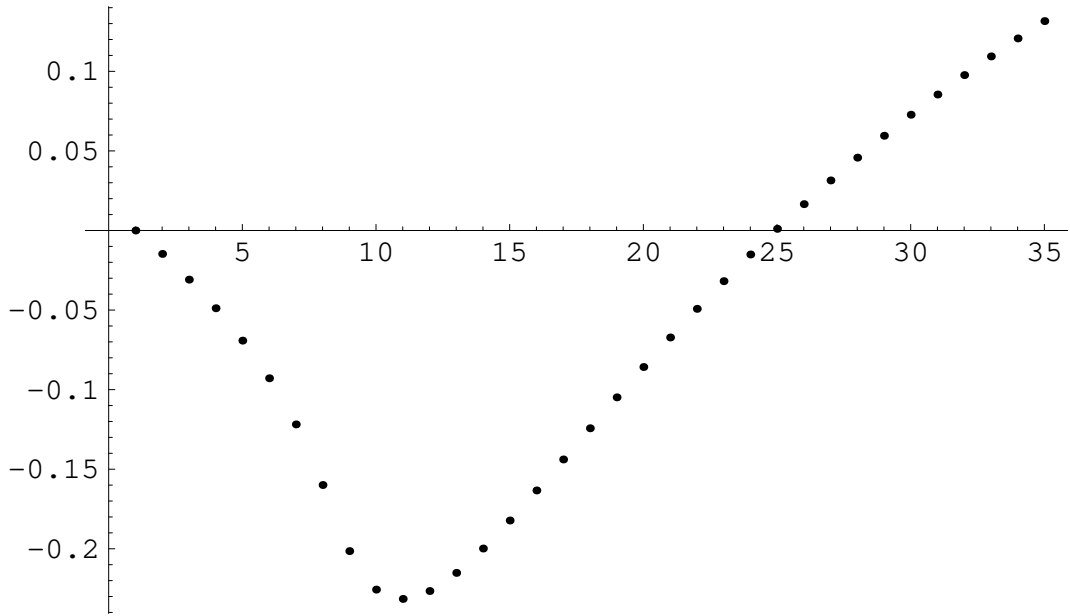}
\caption{Binding energies $B_n=E_n-E_0-nM$ plotted as a function of $n$ for $g=
0.008333$ and $\epsilon=1$. A negative value of $B_n$ indicates a bound state.
Observe that there are 23 bound states for these parameter values. Note that
$B_n$ is a smooth function of $n$.}
\label{f4}
\end{figure}

To display the bound states we simply plot the value of the binding energy $B_n$
as a function of $n$. For example, in Fig.~\ref{f4} we display the bound states
for $\epsilon=1$ and $g=0.008333$. Note that for these values there are 23 bound
states. Observe also that the binding energy $B_n$ is a smooth function of $n$.

\begin{figure}[b!]
\vspace{3.8in}
\includegraphics{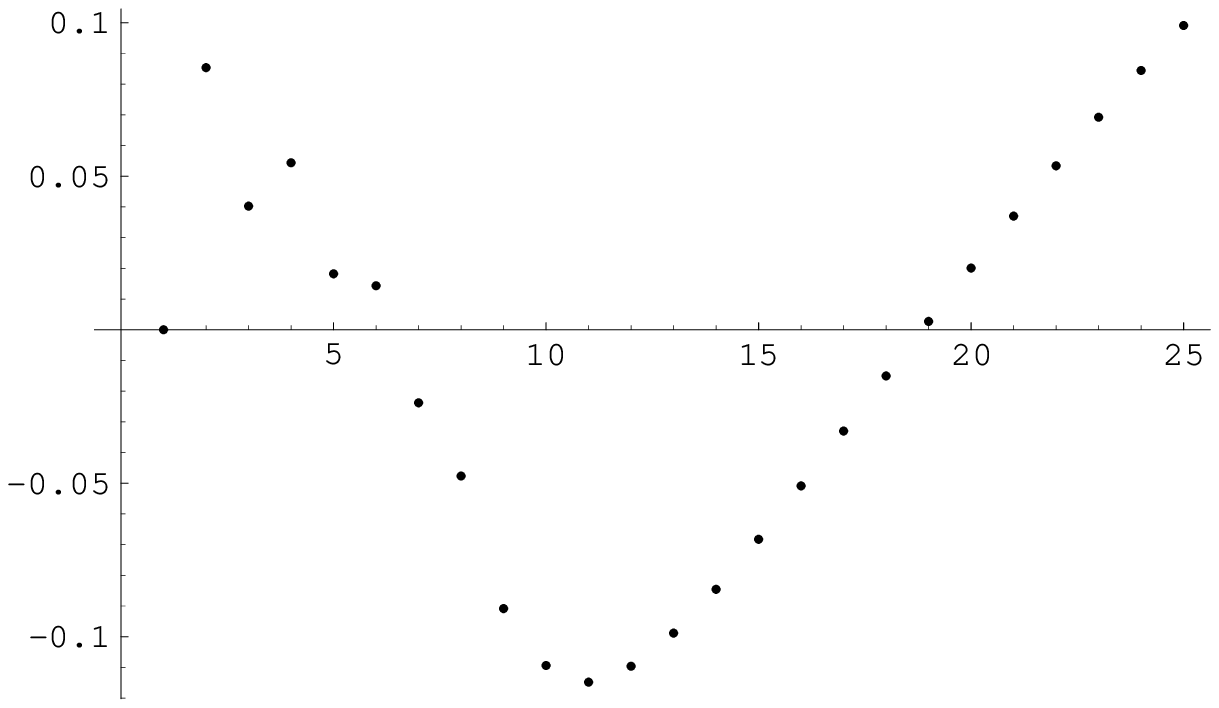}
\caption{Binding energies $B_n$ plotted as a function of $n$ for $g=0.008333$
and $\epsilon=0.9$. Note that the first five states have now become unbound and
$B_n$ is not a smooth function of $n$ for $n\leq6$. The next twelve states are
bound, and in this region $B_n$ is a smooth function of $n$. A comparison of
this figure with Fig.~\ref{f4} shows that the bound-state spectrum is
exquisitely sensitive to the strength of the linear anomaly term.}
\label{f5}
\end{figure}

It is interesting that the bound-state spectrum depends sensitively on the
anomaly term in the Hamiltonian (\ref{x4}). If $\epsilon$ is slightly less
than $1$, the first few states become unbound, as is shown in Fig.~\ref{f5}.
In this figure $g=0.008333$ and $\epsilon=0.9$.

\begin{figure}[b!]
\vspace{3.8in}
\includegraphics{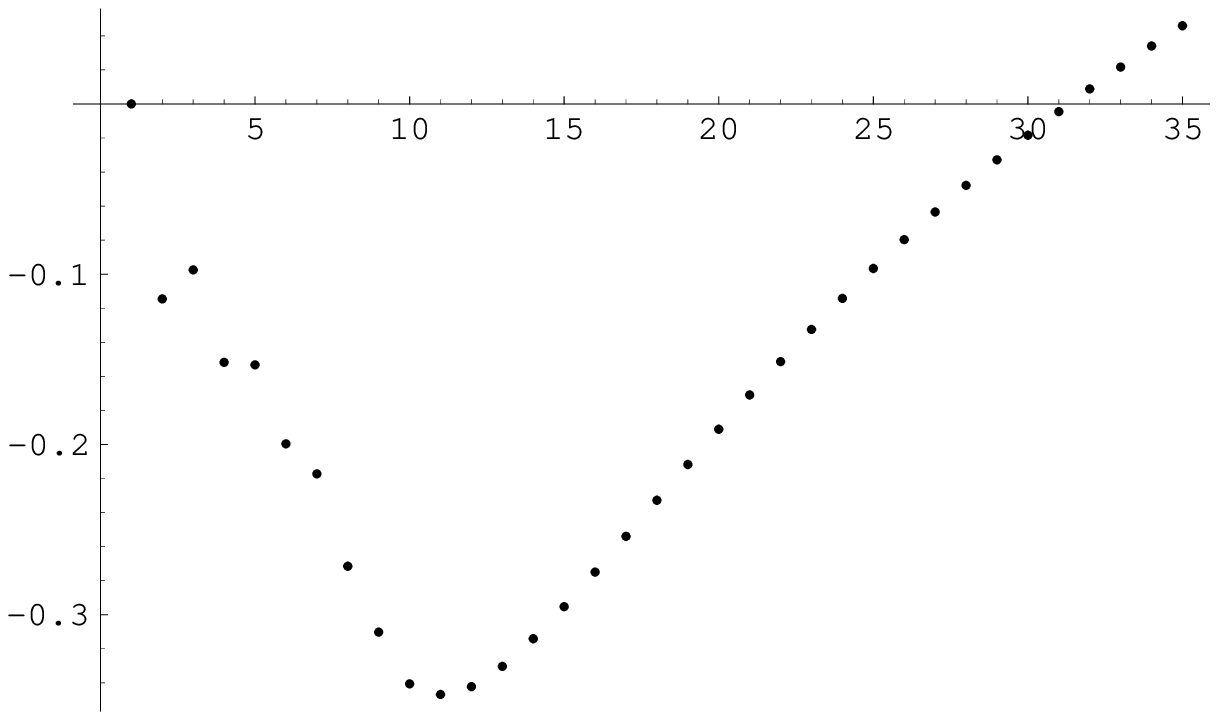}
\caption{Binding energies $B_n$ plotted as a function of $n$ for $g=0.008333$
and $\epsilon=1.1$. Note that there are 30 bound states and that $B_n$ is not a
smooth function of $n$ when $n$ is small.}
\label{f6}
\end{figure}

If $\epsilon$ is slightly greater than $1$, the binding energy $B_n$ is not
a smooth function of $n$ for small $n$. In Fig.~\ref{f6} we plot $B_n$ as a
function of $n$ for $g=0.008333$ and $\epsilon=1.1$. Note that for these
values of the parameters there are 30 bound states. Figures~\ref{f4}, \ref{f5},
and \ref{f6} are strikingly different, and demonstrate the extreme sensitivity
of the bound-state spectrum to the anomaly term.

\section{Path-Integral Derivation of the Parity Anomaly}
\label{s3}
A path integral is used to represent the partition function of a quantum field
theory. A path integral is much more complicated than a Schr\"odinger equation,
and thus the elementary differential-equation methods applied in Sec.~\ref{s2}
cannot be used directly for a quantum field theory. To prepare for our study of
a $\cP\cT$-symmetric $-\phi^4$ field theory in Sec.~\ref{s4}, we need to
establish the equivalence of $H$ in (\ref{e1}) and $\tilde H$ in (\ref{ee14}) by
using path-integration methods.

For the Hamiltonian (\ref{e1}) the Euclidean functional integral for the
partition function $Z$ is
\begin{eqnarray}
Z=\int_C D\phi\,\exp\left[-\frac{1}{\hbar}\int dt\left(\frac{m}{2}\dot\phi^2-
g\phi^4\right)\right],
\label{eee1}
\end{eqnarray}
where the normalization factor is understood. A functional integral is an
infinite product of ordinary integrals, one for each lattice point in the
discretized representation. The $g\phi^4$ term in the exponent would cause each
of these integrals to diverge if the integration path lay on the real axis. To
make these integrals converge, the contour of integration at each lattice point
must approach infinity inside a pair of wedges having an angular opening of
$45^\circ$ and centered about the angles $-45^\circ$ and $-135^\circ$ in the
lower-half
complex plane. The subscript $C$ on the functional integral sign indicates that
the path integral is taken along a complex contour.

Our goal is to transform the functional integral (\ref{eee1}) into a
conventional
functional integral in which the contour of integration runs along the real axis
rather than in the complex plane. We expect to find that the action in the
exponent will correspond to $\tilde H$ in (\ref{ee14}).

Our approach is as follows: First, in Subsec.~\ref{ss1} we transform the
continuum functional integral in (\ref{eee1}) directly using (\ref{ee2}). We
discover that if we proceed formally without recognizing that a functional
integral is a subtle construct involving the limit of a discrete product of
ordinary integrals, the resulting equivalent Hermitian Hamiltonian that we
obtain does not contain the linear anomaly term in $\tilde H$ in (\ref{ee14}).
Thus, this derivation produces the classical ($\hbar\to0$) limit of $\tilde H$.
Hence, this derivation is equivalent to the geometrical-optics approximation
(zeroth-order WKB). To obtain the anomaly, we must discretize the functional
integral, and this discretization requires great care. We explain and motivate
the discretization procedure in Subsec.~\ref{ss2} and carry it out in
Subsec.~\ref{ss3}. 

\subsection{Formal Derivation Correct to Order $\hbar^0$}
\label{ss1}

We begin by making a substitution analogous to that in (\ref{ee2}),
\begin{eqnarray}
\phi(t)=-2iL\sqrt{1+i\psi(t)/L}.
\label{eee2}
\end{eqnarray}
This substitution introduces a functional Jacobian in the form of a square root:
\begin{eqnarray}
D\phi=\frac{D\psi}{{\rm Det}\,\sqrt{1+i\psi/L}}.
\label{eee3}
\end{eqnarray}
The new functional integral over the $\psi$ variable is
\begin{eqnarray}
Z=\int\frac{D\psi}{{\rm Det}\,\sqrt{1+i\psi/L}}
\exp\left\{-\frac{1}{\hbar}\int dt\,\left[\frac{m}{2}\frac{\dot\psi^2(t)}
{1+i\psi(t)/L}-16gL^4\left(1+i\frac{\psi(t)}{L}\right)^2\right]\right\}.
\label{eee4}
\end{eqnarray}
This is a conventional functional integral in the sense that the field $\psi(t)$
is real and the path of integration lies on the real axis rather than in the
complex plane.

Next, we exploit the functional-integral version of the simple integral identity
\begin{eqnarray}
\frac{1}{A}=\frac{1}{\sqrt{2\pi}}\int_{-\infty}^{\infty}dt\,e^{-\half A^2(t-B)^2
}\qquad({\rm Re}\,A^2>0)
\label{eee5}
\end{eqnarray}
and obtain (apart from an overall normalization factor)
\begin{eqnarray}
Z&=&\int D\psi \int D\pi \,\exp\left\{ -\frac{1}{\hbar}\int dt\, \left[
\frac{1+i\psi(t)/L}{2m}\Big(\pi(t)-B(t)\Big)^2\right.\right.\nonumber\\
&&\qquad\qquad\left.\left.
+\frac{m}{2}\frac{\dot\psi^2(t)}{1+i\psi(t)/L}-16gL^4\Big(1+i\psi(t)/L\Big)^2
\right]\right\}.
\label{eee6}
\end{eqnarray}
This integral converges because ${\rm Re}[1+i\psi(t)/L]=1$, which is positive.
Both functional integrals in (\ref{eee6}) are conventional real integrals.

We eliminate the term containing the square of $\dot\psi(t)$ by choosing 
\begin{eqnarray}
B(t)=\frac{im\dot\psi(t)}{1+i\psi(t)/L}.
\label{eee7}
\end{eqnarray}
With this choice the exponential in (\ref{eee6}) simplifies to:
\begin{eqnarray}
-\frac{1}{\hbar}\int dt\left[\frac{1}{2m}\pi^2(t)\Big(1+i\psi(t)/L\Big)
-i\pi(t)\dot\psi(t)-16gL^4\Big(1+i\psi(t)/L\Big)^2\right].
\label{eee8}
\end{eqnarray}

Next, we integrate by parts in order to transfer the derivative from $\psi(t)$
to $\pi(t)$ and in doing so discard the surface term:
\begin{eqnarray}
-\int dt\,\dot\psi(t)\pi(t)=\int dt\,\dot \pi(t)\psi(t).
\label{eee9}
\end{eqnarray}
After interchanging orders of integration and completing the square, we get
\begin{eqnarray}
Z&=&\int D\pi\,\exp\left[-\frac{1}{\hbar}\int dt\left(
\frac{1}{64gL^2}\dot \pi^2(t) +\frac{1}{256gm^2L^4}\pi^4(t) -L\dot \pi(t)
+\frac{1}{64gmL^3}\pi^2(t)\dot \pi(t)\right)\right] \nonumber\\
&&\quad\times \int D\psi\,\exp\left[
-\frac{16gL^2}{\hbar}\int dt\left(\psi(t)-iL+\frac{i}{64gmL^3}\pi^2(t)
+\frac{i}{32gL^2}\dot \pi(t)\right)^2\right].
\label{eee10}
\end{eqnarray}
We integrate the $\dot\pi(t)$ and $\pi^2(t)\dot \pi(t)$ terms and ignore the
surface contributions.

Finally, we rescale $\pi$ and $\psi$ in analogy with (\ref{ee12}):
\begin{eqnarray}
\pi(t)=L\sqrt{32mg}\varphi(t),\qquad\psi(t)=\frac{1}{L\sqrt{32mg}}p(t),
\label{eee11}
\end{eqnarray}
and evaluate the Gaussian integral over $p$. The result is
\begin{eqnarray}
Z=\int D\varphi\,\exp\left[-\frac{1}{\hbar}\int dt\left(
\frac{m}{2}\dot\varphi^2(t)+4g\varphi^4(t)\right)\right],
\label{eee12}
\end{eqnarray}
which is the Euclidean functional integral for the {\it classical\/}
(anomaly-free) version of the Hamiltonian $\tilde H$ in (\ref{ee14}).

\subsection{Discretization of the Function Integral}
\label{ss2}
Evidently, the analysis in Subsec.~\ref{ss1} is not delicate enough to recover
the linear anomaly term of $\tilde H$ in (\ref{ee14}). In order to obtain this
anomaly term it is necessary to discretize the functional integral (\ref{eee1}).
In order to discretize, we replace $\dot\phi(t)$ by the usual lattice expression
$(\phi_{n+1}-\phi_n)/a$, where $a$ is the lattice spacing. Furthermore, we must
replace $\phi(t)$ by the average $\half(\phi_{n+1}+\phi_n)$ to preserve the time
independence of the equal-time commutation relation in the underlying
quantum-mechanical theory. Our purpose in this subsection is to explain the
reason behind this latter substitution.

Consider a one-dimensional quantum-mechanical Hamiltonian of the form 
\begin{eqnarray}
H=\half p^2+V(q).
\label{eee13}
\end{eqnarray}
The Heisenberg equations of motion of the operators $p(t)$ and $q(t)$ have
the same form as the classical equations in (\ref{e3}):
\begin{eqnarray}
\dot x(t)=p,\qquad \dot p(t)=-V'[q(t)].
\label{eee14}
\end{eqnarray}
To discretize the differential equations in (\ref{eee14}), we write
\begin{eqnarray}
\frac{x_{n+1}-x_n}{a}=\frac{p_{n+1}+p_n}{2},\qquad
\frac{p_{n+1}-p_n}{a}=-V'\!\left(\frac{x_{n+1}+x_n}{2}\right),
\label{eee15}
\end{eqnarray}
where $a$ is the lattice spacing.

One can verify that (\ref{eee15}) is correct by following the procedure
introduced in Ref.~\cite{BS}. The right side of the first equation in
(\ref{eee15})
is a function of $p_{n+1}+p_n$, so the commutator of this equation with $p_{n+1}
+p_n$ is
\begin{eqnarray}
\left[x_{n+1}-x_n,p_{n+1}+p_n\right]=0.
\label{eee16}
\end{eqnarray}
Also, the right side of second equation in (\ref{eee15}) is a function of $x_{n+
1}+x_n$. Thus, commuting $x_{n+1}+x_n$ with this equation gives
\begin{eqnarray}
\left[x_{n+1}+x_n,p_{n+1}-p_n\right]=0.
\label{eee17}
\end{eqnarray}
Adding (\ref{eee16}) and (\ref{eee17}) then gives
\begin{eqnarray}
\left[x_{n+1},p_{n+1}\right]=\left[x_n,p_n\right].
\label{eee18}
\end{eqnarray}
This establishes the crucial result that the equal-time commutator is {\it
exactly\/} preserved in time. Thus, the discretization scheme in (\ref{eee15})
is exactly unitary. If a discretization scheme other than that in (\ref{eee15})
had been used, there would be a small violation of unitarity. Evidently,
to avoid violating unitarity on the lattice it is essential to replace a local
function of $x(t)$, say $f[x(t)]$ by $f\!\left[\half\left(x_{n+1}+x_n\right)
\right]$ rather than by $f\!\left(x_n\right)$. Of course, the violation of
unitarity vanishes in the continuum limit $a\to0$, but we need to preserve
unitarity in the lattice version of the theory.

To discretize an action, we follow the same approach as that used to obtain
(\ref{eee15}). Thus, for the Lagrangian $L=\half\dot x^2(t)-V[x(t)]$ we
discretize the action $\int dt\,L$ as follows:
\begin{eqnarray}
\sum_n\left\{\frac{1}{2a}\left(x_{n+1}-x_n\right)^2-aV\!\left[\half\left(x_{n+1}
+x_n\right)\right]\right\}.
\label{eee19}
\end{eqnarray}

To verify that this is the correct way to discretize the action, we vary this
discrete action with respect to $x_n$ to obtain the lattice equations of motion:
\begin{eqnarray}
\frac{1}{a}\left(2x_n-x_{n+1}-x_{n-1}\right)-\frac{a}{2}V'\!\left[\half\left(x_{
n+1}+x_n\right)\right]-\frac{a}{2}V'\!\left[\half\left(x_n+x_{n-1}\right)\right]
=0.
\label{eee20}
\end{eqnarray}
To solve this equation, we substitute
\begin{eqnarray}
\frac{x_{n+1}-x_n}{a}=\frac{p_{n+1}+p_n}{2},
\label{eee21}
\end{eqnarray}
which is the first equation in (\ref{eee15}). This reduces (\ref{eee20}) to
the form 
\begin{eqnarray}
b_n+b_{n-1}=0,
\label{eee22}
\end{eqnarray}
where
\begin{eqnarray}
b_n=\frac{p_{n+1}-p_n}{a}+V'\!\left(\frac{x_{n+1}+x_n}{2}\right).
\label{eee23}
\end{eqnarray}
The solution to (\ref{eee22}) is
\begin{eqnarray}
b_n=c(-1)^n,
\label{eee24}
\end{eqnarray}
where $c$ is an arbitrary constant. However, if $c$ is nonzero, then in the
continuum limit where the lattice spacing tends to $0$ the solution in
(\ref{eee24}) becomes infinitely oscillatory. To avoid the appearance of an
infinite-energy solution, we must require that $c=0$. We therefore obtain the
result that $b_n=0$, and we have recovered the second equation in (\ref{eee15}).
We have reproduced both equations in (\ref{eee15}) and hence conclude that the
discretization scheme used in (\ref{eee19}) is correct. The lattice average
used on the right sides of (\ref{eee15}) and in the second term of (\ref{eee19})
is completely consistent with Moyal ordering \cite{MOYAL}.

\subsection{Derivation of the Parity Anomaly}
\label{ss3}

\begin{figure*}[t!]
\vspace{2.5in}
\includegraphics{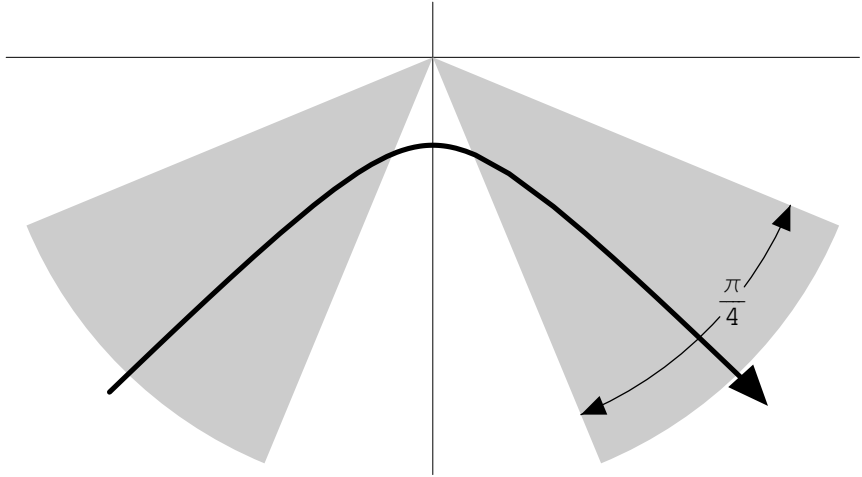}
\caption{Stokes' wedges in the lower-half complex-$\phi_n$ plane for each of the
one-dimensional complex integrals used in (\ref{eee25}). The integrands of these
complex integrals vanish exponentially fast as $|\phi_n|\to\infty$ within these
wedges. Also shown is the complex contour in (\ref{eee26}).}
\label{f7}
\end{figure*}

Deriving the Hamiltonian $\tilde H$ in (\ref{ee14}) from the partition function
$Z$ in (\ref{eee1}) requires great care. First, we must discretize the
functional integral (\ref{eee1}) representing $Z$ by following the procedure in
Subsec.~\ref{ss2}:
\begin{eqnarray}
Z=\prod_n\int_C d\phi_n\,\exp\left\{-\frac{a}{\hbar}\left[\frac{m}{2a^2}\left(
\phi_{n+1}-\phi_n\right)^2-\frac{g}{16}\left(\phi_{n+1}+\phi_n\right)^4\right]
\right\},
\label{eee25}
\end{eqnarray}
where $a$ is the lattice spacing. For each of the one-dimensional integrals in
(\ref{eee25}), the contour $C$ begins in a wedge in the lower complex plane
centered about $-135^\circ$ and terminates in the $\cP\cT$-symmetric wedge
centered about $-45^\circ$ (see Fig.~\ref{f7}).

For each of the complex integrals in (\ref{eee25}) we make a substitution
analogous to that in (\ref{ee2}):
\begin{eqnarray}
\phi_n=-2iL\sqrt{1+i\psi_n/L},
\label{eee26}
\end{eqnarray}
where $L$ is an arbitrary positive constant having dimensions of length. At
each lattice point this substitution introduces a Jacobian in the functional
integral;
\begin{eqnarray}
d\phi_n=\frac{d\psi_n}{\sqrt{1+i\psi_n/L}}.
\label{eee27}
\end{eqnarray}
Thus, an exact transcription of (\ref{eee25}) is
\begin{eqnarray}
Z=\prod_n\int_{-\infty}^{\infty}\frac{d\psi_n}{\sqrt{1+i\psi_n/L}}
\exp\left[-\frac{a}{\hbar}\left(\frac{m}{2}\frac{\dot\psi_n^2}{A_n^2}
-16gL^4A_n^4\right)\right],
\label{eee28}
\end{eqnarray}
where we use the shorthand notation 
\begin{eqnarray}
\dot\psi_n\equiv\frac{1}{a}\left(\psi_{n+1}-\psi_n\right)
\label{eee29}
\end{eqnarray}
and
\begin{eqnarray}
A_n\equiv\frac{1}{2}\left(\sqrt{1+i\psi_{n+1}/L}+\sqrt{1+i\psi_n/L}\right).
\label{eee30}
\end{eqnarray}
Note that in (\ref{eee28}) the path of integration of each of the integrals
lies on the real axis.

We now use the integral identity (\ref{eee5}) to introduce the conjugate
auxiliary field $\pi_n$ at each lattice site $n$. To apply this identity we
choose
\begin{eqnarray}
A^2=\frac{a}{m\hbar}A_n^2,\qquad t=\pi_n,\qquad B=B_n,
\label{eee31}
\end{eqnarray}
where $B_n$ is as yet unspecified. The result is
\begin{eqnarray}
Z&=&\prod_n \int_{-\infty}^{\infty} d\psi_n \int_{-\infty}^{\infty}d\pi_n\sqrt{
\frac{a}{2\pi\hbar m}}\frac{A_n}{\sqrt{1+i\psi_n/L}}\nonumber\\
&&\qquad\qquad\times\exp\left\{-\frac{a}{\hbar}\left[\frac{A_n^2}{2m}\left(\pi_n
-B_n\right)^2+\frac{m}{2}\frac{\dot\psi_n^2}{A_n^2} -16gL^4A_n^4\right]\right\}.
\label{eee32}
\end{eqnarray}
Making no approximations, we simplify the Jacobian in (\ref{eee32}) as follows:
\begin{eqnarray}
\frac{A_n}{\sqrt{1+i\psi_n/L}}&=&
\frac{\sqrt{1+i\psi_{n+1}/L}+\sqrt{1+i\psi_n/L}}{2\sqrt{1+i\psi_n/L}}\nonumber\\
&=&1+\frac{\sqrt{1+i\psi_{n+1}/L}-\sqrt{1+i\psi_n/L}}{2\sqrt{1+i\psi_n/L}}
\nonumber\\
&=&1+\frac{i(\psi_{n+1}-\psi_n)}{4LA_n \sqrt{1+i\psi_n/L}}\nonumber\\
&=& 1+ a\frac{i\dot\psi_n}{4LA_n\sqrt{1+i\psi_n/L}}.
\label{eee33}
\end{eqnarray}

Next, we choose $B_n$:
\begin{eqnarray}
B_n=\frac{im\dot\psi_n}{A_n^2}.
\label{eee34}
\end{eqnarray}
[Note that the choice of sign on the right side of (\ref{eee34}) is arbitrary.
If we replace $i$ by $-i$ in this equation, then at the end of the calculation
the sign of the linear term in the Hermitian Hamiltonian will be reversed.
Reversing this sign has no effect on the energy levels of the Hamiltonian.]
With this choice, we can approximate for small lattice spacing $a$ the Jacobian
factor in the last line of (\ref{eee33}):
\begin{eqnarray}
1+a\frac{i\dot\psi_n}{4LA_n\sqrt{1+i\psi_n/L}}=
1+\frac{aB_n}{4mL}+{\rm O}\!\left(a^2\right).
\label{eee35}
\end{eqnarray}

Next, we use the identity
\begin{eqnarray}
\int_{-\infty}^\infty d\pi_n\,\pi_n\,\exp\left[-\frac{aA_n^2}{2m\hbar}\left(
\pi_n-B_n\right)^2\right]=
\int_{-\infty}^\infty d\pi_n\,B_n\,\exp\left[-\frac{aA_n^2}{2m\hbar}\left(
\pi_n-B_n\right)^2\right],
\label{eee36}
\end{eqnarray}
which holds because the exponent is an even function of $\pi_n-B_n$.
This identity implies that (\ref{eee32}) with the Jacobian in (\ref{eee35})
can be rewritten as
\begin{eqnarray}
Z&=&\prod_n \int_{-\infty}^{\infty} d\psi_n \int_{-\infty}^{\infty}d\pi_n\sqrt{
\frac{a}{2\pi\hbar m}}
\left[1+\frac{a\pi_n}{4mL}+{\rm O}\!\left(a^2\right)\right]\nonumber\\
&&\qquad\qquad\times\exp\left[-\frac{a}{\hbar}
\left(\frac{A_n^2\pi_n^2}{2m}-i\pi_n\dot\psi_n-16gL^4A_n^4\right)\right].
\label{eee37}
\end{eqnarray}

Next we promote the Jacobian factor to the exponent by noting that for any
number $w$, we have $1+aw=e^{aw}$ with an error of order $a^2$. We also simplify
terms in the exponent by ignoring terms of order $a^2$. The result is
\begin{eqnarray}
Z&=&\prod_n\int_{-\infty}^{\infty}d\psi_n\int_{-\infty}^{\infty}d\pi_n\sqrt{
\frac{a}{2\pi\hbar m}}\exp\left\{
-\frac{a}{\hbar}\left[\left(1+i\psi_n/L\right)\frac{\pi_n^2}{2m}
+i\dot\pi_n\psi_n\right.\right.\nonumber\\
&&\qquad\qquad\left.\left.  -16gL^4\left(1+i\psi_n/L\right)^2
-\frac{\hbar\pi_n}{4mL}+{\rm O}(a)\right]\right\},
\label{eee38}
\end{eqnarray}
where we have performed a summation by parts to obtain the
$i\dot\pi_n\psi_n$ term.

Next, we perform a scaling like that in (\ref{eee11}):
\begin{eqnarray}
\pi_n=\sqrt{32mg}L\varphi_n,\qquad \psi_n=\frac{1}{\sqrt{32mg}L}p_n,
\label{eee39}
\end{eqnarray}
and continue as we did in Subsec.~\ref{ss1}. After completing the square and
integrating over $p_n$, we obtain the final result for the lattice version of
the continuum functional integral:
\begin{eqnarray}
Z=\int D\varphi\,\exp\left[-\frac{1}{\hbar}\int dt\left(\frac{m}{2}\dot\varphi^2
+4g\varphi^4-\hbar\sqrt{\frac{2g}{m}}\varphi \right)\right].
\label{eee42}
\end{eqnarray}
This is precisely the functional integral for the theory described by the
Hermitian Hamiltonian $\tilde H$ in (\ref{ee14}).

\section{Generalization to Quantum Field Theory}
\label{s4}

The Euclidean functional integral for the partition function $Z$ of a
$d$-dimensional $-\phi^4$ quantum field theory is
\begin{eqnarray}
Z=\int_C D\phi\,\exp\left[-\int d^dx\left(\half(\nabla\phi)^2-g\phi^4\right)
\right],
\label{eeee1}
\end{eqnarray}
where the normalization factor is understood and we have adopted natural
units where $\hbar=1$. As in the one-dimensional quantum field theory in
(\ref{eee1}), the contour of integration at each space-time point must approach
infinity inside a pair of wedges having an angular opening of $45^\circ$ and
centered about the angles $-45^\circ$ and $-135^\circ$ in the lower-half complex
plane. The subscript $C$ on the functional integral sign indicates that the path
integral is taken along a complex contour.

As in the quantum-mechanical case, our goal is to transform the functional
integral (\ref{eeee1}) into a conventional functional integral in which the
contour of integration runs along the real axis rather than in the complex
plane. Following the procedures in Sec.~\ref{s3} we make the substitution
\begin{eqnarray}
\phi(x)=-2i\sqrt{1+i\psi(x)},
\label{eeee2}
\end{eqnarray}
where we have set $L=1$. This substitution introduces a functional Jacobian in
the form of a square root:
\begin{eqnarray}
D\phi=\frac{D\psi}{{\rm Det}\,\sqrt{1+i\psi}}.
\label{eeee3}
\end{eqnarray}
The new functional integral over the $\psi$ variable is
\begin{eqnarray}
Z=\int\frac{D\psi}{{\rm Det}\,\sqrt{1+i\psi}}\exp\left\{-\int d^dx\,\left[\frac{
(\nabla\psi)^2}{2(1+i\psi)}-16g(1+i\psi)^2\right]\right\}.
\label{eeee4}
\end{eqnarray}
This is a conventional functional integral in the sense that the field $\psi(x)$
is real and the path of integration lies on the real axis rather than in the
complex plane.

From the form of the path integral in (\ref{eeee4}) we can see that even though
the complex number $i$ appears in the integrand, $Z$ is real because the change
of variable $\psi\to-\psi$ changes the sign of $i$. Thus, the ground-state
energy density is real.

This same argument implies that the $2n$-point Green's functions are real
and the $(2n+1)$-point Green's functions are imaginary. The $n$-point Green's
function $G_n(x_1,x_2,x_3,\dots,x_n)$ is constructed from the moments of
products of $\phi$ fields:
\begin{eqnarray}
\langle0|\phi(x_1)\phi(x_2)\phi(x_3)\cdots\phi(x_n)|0\rangle&=&
\int_C D\phi\,\phi(x_1)\phi(x_2)\phi(x_3)\cdots\phi(x_n)\nonumber\\
&&\qquad\times\exp\left[-\int d^dx\left(\half(\nabla\phi)^2-g\phi^4\right)
\right].
\label{eeee5}
\end{eqnarray}
After making the substitution (\ref{eeee2}), we can apply the above symmetry
argument to show that $G_n$ is real when $n$ is even and imaginary when $n$
is odd.

Even though we have shown that the partition function $Z$ in (\ref{eeee4}) is
real, we are still unable to identify a Hermitian field-theoretic Lagrangian 
that is equivalent to the $-\phi^4$ non-Hermitian Lagrangian. This is because
the functional integral (\ref{eeee4}) has a Jacobian factor, and this factor is 
complex. By following the procedure in Sec.~\ref{s3}, we were able to eliminate
this Jacobian factor for the case of a one-dimensional quantum field theory. It
is not obvious how to proceed in the case of higher dimensions. We now propose
three possible approaches to solving this problem.

\begin{enumerate}
\item
The simplest approach is to follow Sec.~\ref{s3} and introduce a new field $\pi
(x)$ by using the identity (\ref{eee5}). The advantage of this procedure is
that, at least at the classical level where we do not perform the
point-splitting described in Subsec.~\ref{ss3}, the Jacobian factor in the
functional integral cancels. We obtain the $d$-dimensional analog of
(\ref{eee6}):
\begin{eqnarray}
Z=\int\!D\psi\int\!D\pi\,\exp\left\{-\!\int\!d^dx\left[\frac{1+i\psi}{2}
(\pi-B)^2+\frac{(\nabla\psi)^2}{2(1+i\psi)}-16g(1+i\psi)^2\right]\right\}.
\label{eeee6}
\end{eqnarray}
There are now several possibilities for the choice of $B$. The choice
\begin{eqnarray}
B(x)=\frac{i\nabla_1\psi(x)}{1+i\psi(x)}
\label{eeee7}
\end{eqnarray}
eliminates the $[\nabla_1\psi(x)]^2$ term from (\ref{eeee6}) but the resulting
action is nonpolynomial in $\psi$ and thus the integral over $\psi$ cannot
be evaluated analytically. Furthermore, the resulting action is not manifestly
Lorentz covariant. The more symmetric choice 
\begin{eqnarray}
B(x)=\frac{i\sqrt{(\nabla\psi)^2}}{1+i\psi(x)}
\label{eeee8}
\end{eqnarray}
eliminates the $(\nabla\psi)^2/(1+i\psi)$ term completely, but again the
resulting action is nonpolynomial and it is impossible to perform the $\psi$
integral.

\item
A second approach is to introduce the $d$-dimensional vector field $\pi_\mu$
$(\mu=1,\dots d)$. Now, using the identity (\ref{eee5}) $d$ times, we obtain an
action that is a quadratic polynomial in $\psi$. However, we still cannot
perform the $\psi$ integration because the Jacobian factor no longer cancels.
Indeed, there are now $d-1$ functional determinant factors of $\sqrt{1+i\psi}$
in the numerator. Furthermore, had we kept the scale factor $L$ in the change of
variables (\ref{eeee2}), the resulting Jacobian would have explicit $L$
dependence, even though the partition function is clearly independent of $L$.
Therefore, this integral representation of the partition function is clumsy
and is likely to be intractable.

\item
We believe that the most promising approach is to introduce a single scalar
field $\rho(x)$ by formally rewriting $Z$ in (\ref{eeee4}) as
\begin{eqnarray}
Z&=&\int D\psi\frac{{\rm Det}\,\sqrt{\nabla_\mu(1+i\psi)\nabla_\mu}}
{{\rm Det}\,\sqrt{1+i\psi}}\nonumber\\
&&\times\int D\rho\,
\exp\left\{-\int d^dx\,\left[\half(\nabla\rho)^2(1+i\psi)-i\nabla_\mu\rho
\nabla_\mu\psi-16g(1+i\psi)^2\right]\right\}.
\label{eeee9}
\end{eqnarray}
The advantage of this representation of $Z$ is that the action is quadratic
in $\psi$ and is Lorentz covariant. The Jacobian remains complicated, but it
may become analytically tractable if we introduce a Grassmann integration
variable $\eta$ to promote the functional determinant to the action. After
integrating out the $\psi$ field, the action will now depend on the scalar
boson field $\rho$ and the fermion field $\eta$. The resulting Hermitian action
will bear a strong resemblance to supersymmetric actions. This is not surprising
because, even in the one-dimensional case, the equivalent Hermitian Hamiltonian
$\tilde H$ in (\ref{ee14}) strongly resembles a supersymmetric quantum theory.
In a future paper we hope to present a thorough analysis of this quantum theory.

\end{enumerate}

\vskip1pc KAM thanks the Physics Department at Washington University for its
hospitality. We are grateful to the U.S.~Department of Energy for financial
support.

\end{document}